\documentclass[12pt,osajnl2,preprint,showpacs]{revtex4}

\usepackage{epsfig}
\usepackage{graphicx}  %must use pdf pictures if going from latex to pdf straight away.
\usepackage{dcolumn}% Align table columns on decimal point
\usepackage{bm}% bold math
\usepackage{color}% color text
\usepackage{hyperref} %%for creating hyperlink
\begin{document}
\DeclareRobustCommand{\baselinestretch{2.2}}
\title{Nonlinear Femtosecond Pulse Reshaping in Waveguide Arrays
%in normally dispersive fiber
%\\ lasers using waveguide arrays
}
\author{Darren D. Hudson$^{1}$$^{*}$, Kimberlee Shish$^{2}$,
Thomas R. Schibli$^{1}$, J. Nathan Kutz$^{2}$, Demetrios N. Christodoulides$^{3}$, Roberto Morandotti$^{4}$,
and Steven T. Cundiff$^{1}$}

\affiliation{
$^{1}$ JILA, National Institute of Standards and Technology and University of Colorado, Boulder, CO 80309-0440 \\
$^{2}$ Department of Applied Mathematics, University of Washington, Seattle,
WA  98195-2420\\
$^{3}$ College of Optics and Photonics, CREOL, University of Central Florida, 4000 Central Florida Boulevard, Orlando, FL  32816\\
$^{4}$ Institut National de la Recherche Scientifique, Universit\'e du Quebec, 1650 Boulevard Lionel Boulet, J3X 1S2, Varennes, Canada\\
$^{*}$ Department of Physics, University of Colorado, Boulder, CO  80309-0390
 }
\date{\today}
\begin{abstract}
We observe nonlinear pulse reshaping of femtosecond pulses in a waveguide array
due to coupling between waveguides.  Amplified pulses from a mode-locked fiber
laser are coupled to an AlGaAs core waveguide array structure.  The observed
power-dependent pulse reshaping agrees with theory, including shortening of the
pulse in the central waveguide.
\end{abstract}
\maketitle

Waveguide arrays have proven to be an ideal testbed for nonlinear optical waves
in discrete systems.  Much of the work done on these devices has focused on the
spatial aspects of the output light
field~\cite{array9,array1,MI,SurfaceSolitons}.  The physics that determines the
spatial profile of the transmitted beam is quite rich and suitable to
theoretical treatment via the discrete nonlinear Schr\"odinger equation.  This
nonlinear equation can be tailored to include many physical processes that are
present in the waveguide array such as discrete diffraction, normal dispersion,
and self-phase modulation.  Many novel spatial phenomena have been demonstrated
using a waveguide array including discrete spatial
solitons~\cite{array9,array1}, discrete modulational instability~\cite{MI}, and
optical discrete surface solitons~\cite{SurfaceSolitons}.

Despite the fact that most of these experiments used pulses to achieve the
necessary peak powers, temporal effects have largely been neglected.  Recently,
the formation of X-waves was demonstrated in waveguide
arrays~\cite{array7,Droulias,KutzXwave}.  The general structure observed was
attributed to the interplay between discrete diffraction, normal dispersion, and
the nonlinear Kerr effect.  Here, we demonstrate nonlinear pulse shortening in a
waveguide array.  We carefully examine how the pulse shape in each waveguide
depends on peak power of the input pulse using intensity autocorrelation.  The
output of the central waveguide shows significant shortening for high peak power
due to attenuation of its lower power wings, as predicted
recently~\cite{array6}.  The energy in the wings shifts to the outer waveguides.
Thus, the waveguide array is acting as an effective saturable absorber.
Understanding the temporal reshaping in the waveguide will be critical for
future work involving these devices, such as using the waveguide array as a
pulse reshaper in optical telecommunication systems or as an intensity
discriminator inside a mode-locked laser~\cite{array10}.  Simulations of the
governing coupled-mode equations corroborate the observed experimental
pulse-shaping results.

Coupled-mode theory provides an analytic reduction of the governing equations
describing the propagation of electromagnetic energy in waveguides and waveguide
arrays~\cite{marcuse}.  The theory assumes that the electromagnetic field is
localized transversely in each waveguide and that the exchange of energy between
the waveguides can be accurately modeled by an evanescent, linear coupling. When
intense electromagnetic fields induce self-phase modulation, coupled-mode theory
can be modified to include the nonlinear index of refraction~\cite{array9}. The
resulting nonlinear coupled-mode theory agrees well with
experiment~\cite{array1,array2,array3,array4}.  To leading-order, the
nearest-neighbor coupling of electromagnetic energy in the waveguide array is
included in the discretely coupled nonlinear Schr\"odinger equations

\begin{equation}
  i \frac{\partial A_n}{\partial z} - \frac{\beta''}{2} \frac{\partial^2 A_n}{\partial t^2}
   + \gamma |A_n|^2 A_n + c (A_{n+1} + A_{n-1} )  = 0.
   \label{eq:NLSE}
\end{equation}

where $A_n$ represents the normalized electric field amplitude in the $n^{th}$ waveguide ($n=-N,\cdots , -1, 0, 1, \cdots ,N$ and there are $2N+1$ waveguides).  We take the linear  coupling coefficient to be $c=0.82$~mm$^{-1}$ and the
nonlinear self-phase modulation parameter to be $\gamma=3.6$~m$^{-1}$W$^{-1}$. The parameter $\beta''=1.25$ ps$^{2}$/m is the experimentally measured chromatic dispersion in the waveguide array. The simulations of Eq.~(\ref{eq:NLSE}) that follow are performed with 41 ($N=20$) waveguides~\cite{array4} for various launch powers that match experimental conditions.  A pseudo-spectral method is implemented that spectrally transforms the time-domain solution and uses a fourth-order Runge-Kutta for propagation in the waveguide.

\begin{figure}
% use one of the following
%%\includegraphics[width=7.5cm]{figure/figname.eps} % file in subdir figure
\includegraphics[width=8.5cm]{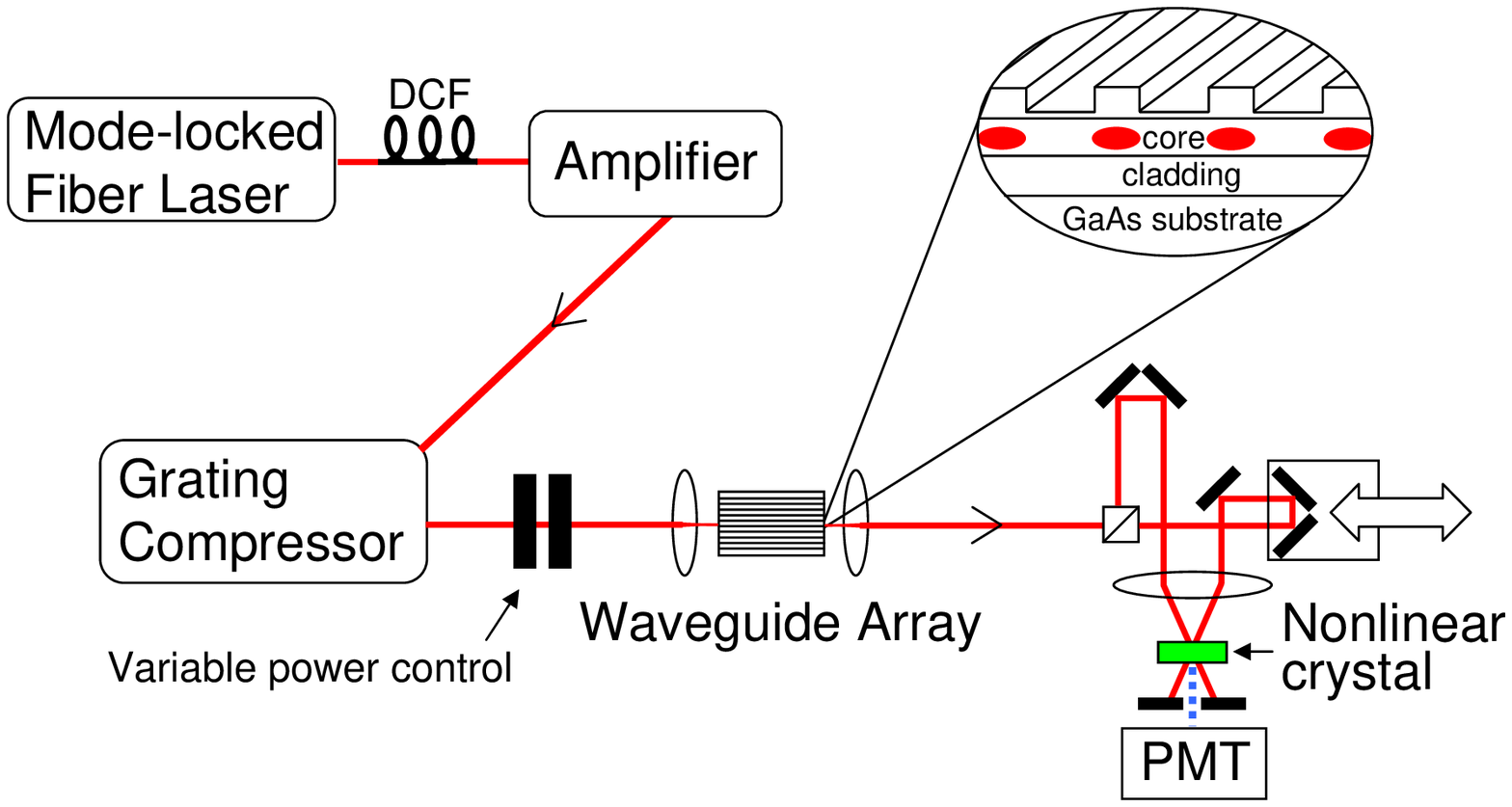} % file in same dir as tex file
\caption{\label{fig:setup}}
\end{figure}

To generate the input pulses, we use a mode-locked, Erbium doped fiber laser
with a repetition rate of 25 MHz (operating at 1550nm) and a chirped-pulse
amplifier/compressor system (see Fig.~\ref{fig:setup}).  Using dispersion
compensating fiber (DCF), the normally chirped pulses from the fiber laser are
further broadened to several picoseconds to avoid nonlinearities in the
amplifier.  These stretched pulses are coupled to a bi-directionally pumped
Erbium amplifier~\cite{fiberAmplifier}, which increases the pulse energy by a
factor of 7, while maintaining the original pulse shape.  The output of the
amplifier is temporally compressed/stretched in free-space by a diffraction
grating compressor.  The compressor is adjusted to produce 600 fs pulses (FWHM
as measured by autocorrelation), which are normally chirped and 3.8 times the
Fourier transform limit.  The output pulse energy is 3.5 nJ.

The pulses are coupled into the waveguide array using standard microscope
objectives (40x) mounted on 3-axis stages.  The input field is mode matched to
the waveguide with a coupling efficiency $>$ 60\%, corresponding to a peak power
of 1.5 kW.  The waveguide array has a 10 $\mu$m center-to-center spacing between
waveguides, with 1.5 $\mu$m tall ridges and 4 $\mu$m wide waveguides.  Index
guiding in the vertical direction is provided by a core layer consisting of
Al$_{0.18}$Ga$_{.82}$As and cladding layers consisting of
Al$_{0.24}$Ga$_{.76}$As.

\begin{figure}
\vspace{-1.2cm}
\includegraphics[width=7.5cm]{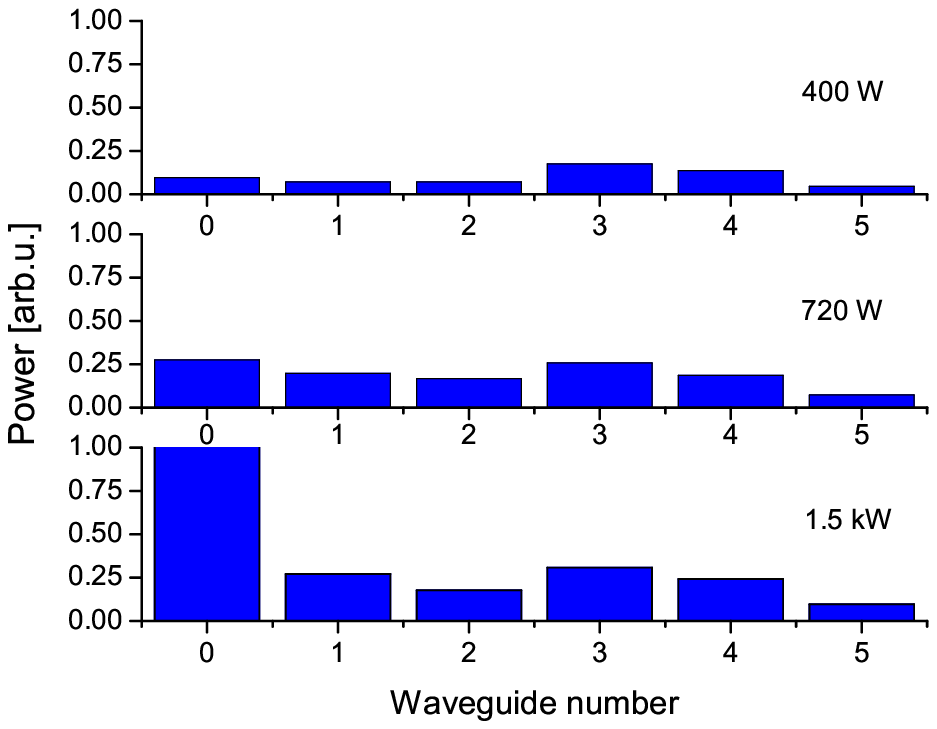}
\vspace{-.4cm}
\caption{\label{fig:PowerDistribution}}
\end{figure}

To verify that discrete spatial solitons are forming and to estimate the
coupling coefficient between adjacent waveguides, the output power distribution
of the array was measured as a function of input power (see
Fig.~\ref{fig:PowerDistribution}).  The energy localizes in the center
waveguides for high power due to discrete self-focusing in the waveguide
array~\cite{array1}.  At low power, the input light easily couples to
neighboring waveguides and thus yields a nearly uniform power in each waveguide
at the output end.

To measure the temporal reshaping effects of the waveguide array, background
free autocorrelations are performed on the output of each waveguide.  The
autocorrelation measurements are performed in the crossed-beam geometry with a
thin BBO crystal used for Type-1 second harmonic generation (SHG).  A
translation stage provides a scanning delay, while a 16-bit digitizer records
the SHG signal detected by a photomultiplier tube.  The data traces are
continuously scanned and averaged.  For reference, an autocorrelation of the
input pulse is also recorded.

\begin{figure}
\includegraphics[width=7.5cm]{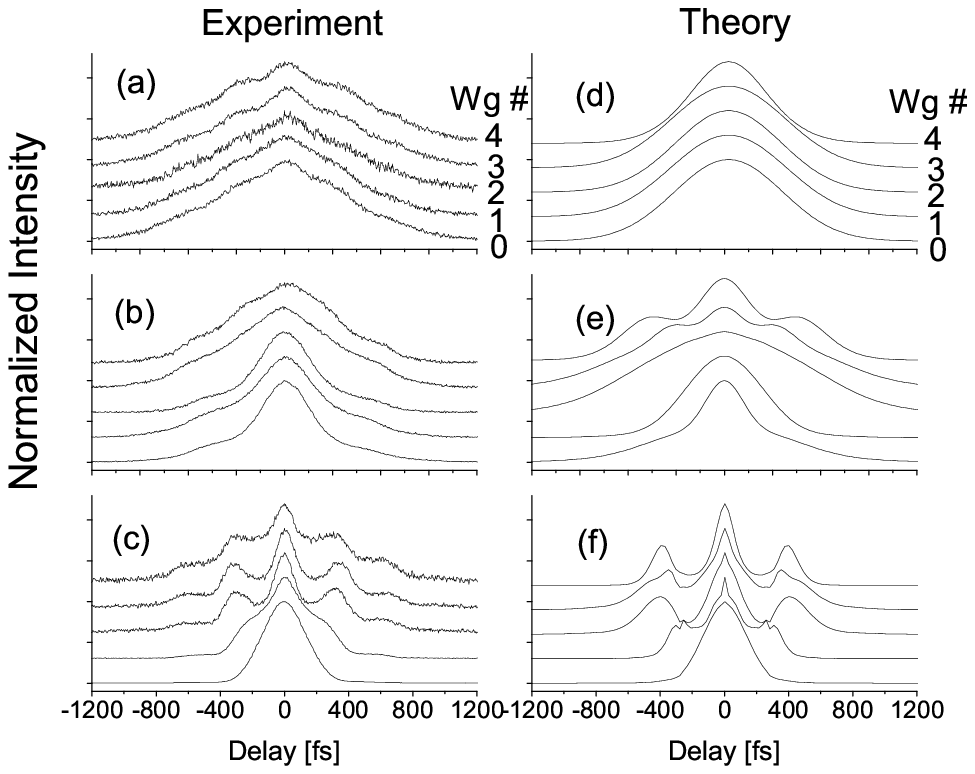} % file in same dir as tex file
\vspace{-0.2cm}
\caption{\label{fig:ShapeVWg}}
\end{figure}

Fig.~\ref{fig:ShapeVWg} shows the pulse reshaping effects of the waveguide array
at each of the input powers, with experimental results on the left and numerical
simulation of Eq.~(\ref{eq:NLSE}) on the right.  At a peak power of 400 W, the
output pulses from the central and outer waveguides were essentially identical
to the input pulse (Fig~\ref{fig:ShapeVWg}-(a) and (d)).  In this regime the
$\gamma$ term is neglible.  The weak pulse launched into the center waveguide
evanescently couples to neighboring waveguides.  Thus, at the output multiple
copies of the input pulse can be observed in each waveguide.  As the input power
is increased to 720 W ((b) and (e)), the pulse reshaping of the central
waveguide begins to emerge.  At the highest input power (1.5 kW) the $\gamma$
term in Eq.~(\ref{eq:NLSE}) becomes non-negligible and the peak of the pulse
decouples from neighboring waveguides.  Meanwhile, the low intensity wings of
the pulse are coupled to the nearest neighbor waveguides.  The result is a
shortened pulse in the center waveguide with the wings removed in agreement with
the predicted nonlinear pulse shortening~\cite{array6}.
Fig.~\ref{fig:ShapeVWg}-(c),(f) shows the output of the waveguide array at high
power.  The triple peaked autocorrelation of the outer waveguides in
Fig.~\ref{fig:ShapeVWg}-(c),(f) is evidence of a double peaked pulse shape.  The
experimental results agree well with the numerical simulation at each power
level.

\begin{figure}
\includegraphics[width=7.5cm]{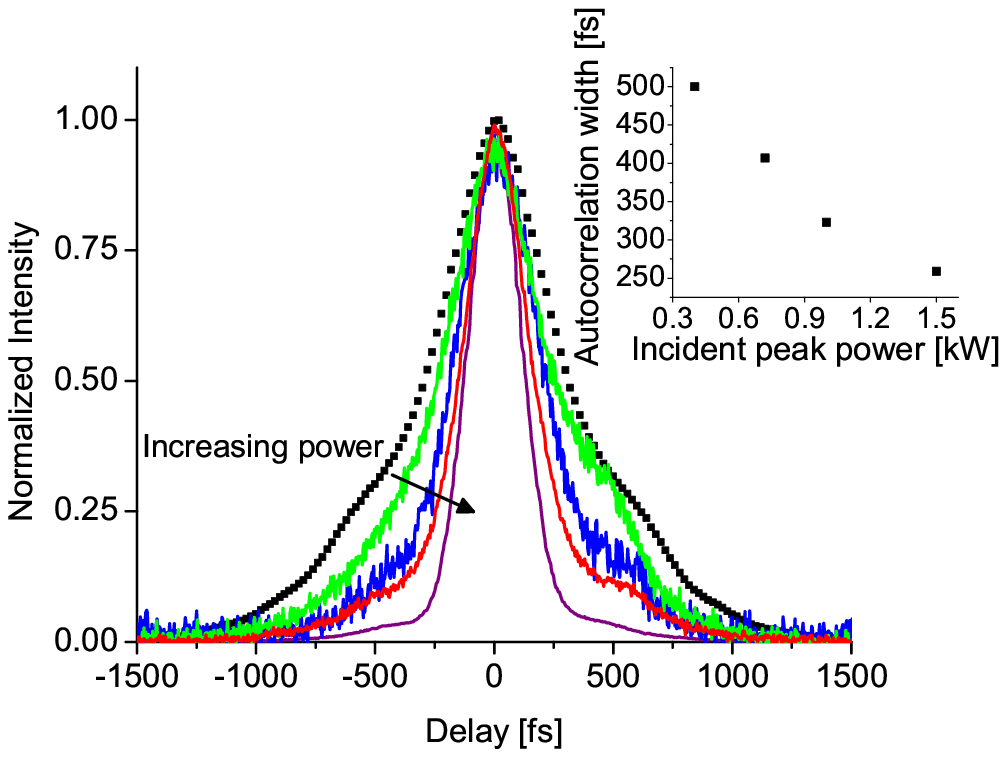}
\vspace{-0.2cm}
\caption{\label{fig:CentralWGshaping}}
\end{figure}

Taking a closer look at the central waveguide pulse shape as a function of input
power (Fig.~\ref{fig:CentralWGshaping}) shows the reshaping increases strongly
at high peak power.  In Fig.~\ref{fig:CentralWGshaping}, a 600 fs pulse was
launched into the central waveguide and the output autocorrelation of the
central waveguide was measured as a function of input power.  The output pulse
for the highest power case shows a pulse width of less than half that of the
input pulse.  To determine how much of the pulse reshaping could be due to
dispersive recompression of the pulse in the waveguide, the dispersion of the
waveguide was measured to be around 1250 fs$^2$/mm using white-light
intereferometry.  Given the length of the waveguide array, dispersion should
only change the pulse length by around 60 fs, well below the change observed
($>$300 fs).  Also, a purely chromatic dispersion compression~\cite{array6}
would be independent of the peak power in the waveguide.

In summary, we have observed nonlinear pulse shortening in a waveguide array and
theoretically matched the experimental results by numerically solving
Eq.~(\ref{eq:NLSE}).  This phenomenon could have a wide range of applications
involving pulse reshaping for long distance telecommunications and emerging
photonic technologies.  In addition, understanding this pulse reshaping will be
crucial for any mode-locked laser system using waveguide array technology as an
effective saturable absorber.  Future work will include a detailed measurement
of the pulse reshaping as a function of the coupling coefficient between
waveguides.

J. N. Kutz acknowledges support from the National Science
Foundation(DMS-0604700).  The authors would also like to thank M.M. Feng and
J.K. Wahlstrand for technical assistance, and A. D. Bristow for useful
discussions.

\clearpage

Fig 1. (color online) Experimental setup.  The output of the fiber laser is
broadened by dispersion compensating fiber (DCF) to avoid nonlinearities in the
amplifer.  The grating compressor is tuned to produce a 600 fs pulse.  The
variable power control consists of a half-wave plate and a polarizer.  A
temporal intensity autcorrelation of the output pulses is recorded using a
photomultiplier tube (PMT).

Fig 2. (color online) Measured power distribution at the output of the waveguide
array. At low peak power the pulse in the center waveguide couples to
neighboring waveguides.  At high peak power, the pulse self-focuses in the
central waveguide which results in a localized power distribution. The waveguide
modes located symmetrically about the central waveguide had a symmetrical power
distribution (not shown).

Fig 3. (color online) Autocorrelation signal versus waveguide number, with
experimental results on the left ((a)-(c)) and theoretical simulations on the
right ((d)-(f)).  The three power levels shown correspond to those in
Fig.~\ref{fig:PowerDistribution}, with (a) and (d) at 400 W, (b) and (e) at 720
W, and (c) and (f) at 1.5 kW.  Pulse shortening in the center waveguide is
observed in the 720 W and 1.5 kW cases.  The autocorrelations are offset
vertically for clarity, with the central waveguide being the lowest and the
outer waveguides plotted sequentially higher on the vertical scale.

Fig 4. (color online) Output autocorrelation of central waveguide for input
powers of 400 W, 720 W, 1 kW, and 1.5 kW.  The inset shows the autocorrelation
FWHM as a function of input power.  The dotted trace is an autocorrelation of
the input pulse.

\clearpage

%%%%%%%%%% Figure 1 %%%%%%%%%%%%
%%%%%%%%%%%%%%%%%%%%%%%%%%%
\clearpage
%%%%%%%%%%Figure 2%%%%%%%%%%%%%%%%
%%%%%%%%%%%%%%%%%%%%%%%%%%%%%%%%%%%
\clearpage
%%%%%%%%%%%Figure 3%%%%%%%%%%%%%%%
%%%%%%%%%%%%%%%%%%%%%%%%%%%%%%%%%%%
\clearpage
%%%%%%%%%%%%Figure 4%%%%%%%%%%%%%%%
%%%%%%%%%%%%%%%%%%%%%%%%%%%%%%%%%%%

%--------------------------------------------------------
\end{document}